\documentclass{he_symp}
\usepackage{psfig}
\begin{document}
\title{Two-stream Instability in Pulsar Magnetospheres}
\author{V.V. Usov}  
\institute{Department of Condensed Matter Physics, Weizmann Institute 
of Science, Rehovot 76100, Israel}
\maketitle

\begin{abstract}
Creation of electron-positron pairs near the pulsar surface 
and the parameters of plasma in pulsar magnetospheres are discussed. 
It is argued that the pair creation process is nonstationary, and
the pair plasma that flows out from the pulsar environment is strongly
nonhomogeneous and gathers into separate clouds. Plasma instabilities 
in the outflowing plasma are reviewed. The two-stream instability 
that develops due to strong nonhomogeneity of the outflowing plasma
is the most plausible reason for the generation of
coherent radio emission of pulsars. The development of the two-stream
instability in pulsar magnetospheres is considered. 
\end{abstract}

\section{Introduction}

Current models for the generation of coherent radio emission of pulsars
require the development of plasma instabilities in the pulsar magnetospheres,
the most studied of which is the two-stream instability 
(for a review, see Arons 1981a,b; Melrose 1981, 1993, 1995;
Usov 1981; Asseo 1996). The two-stream instability in the
pulsar magnetospheres has been considered in many papers (e.g.,
Ruderman \& Sutherland 1975; Benford \& Buschauer 1977; Buschauer \&
Benford 1977; Cheng \& Ruderman 1977; Arons 1981; Usov 1987; Ursov 
\& Usov 1988; Asseo 1993; Asseo \& Melikidze 1998; Melikidze, Gil \&
Pataraya 2000). 

The two-stream instability can arise when two
plasma flows travel through each other: 

$$
\displaylines{\,\,\,\,\,\,\,\,\,\,\,\,\,\,\,\,\,\,
\Longleftarrow\,\,\,{\bf v}_1,\, n_1\cr
{\bf v}_2,\,n_2\,\,\,\Longrightarrow \,\,\,\,\,\,\,\,\,\,\,\,\,
\,\,\,\,\,\,\,\,\,\,\cr}
$$

\noindent
If the density of one flow is much higher than the density of the 
other (for example, $n_1\gg n_2$),
the low-density flow (component~2) calls the beam while the high-density
flow (component 1) calls the plasma. In the magnetospheres of pulsars, both 
the plasma-beam interaction and the interaction of two plasmas 
with more or less equal densities may occur (see below).
The two-stream instability may lead to formation of 
plasma bunches that generate the radio emission via curvature mechanism
(Ruderman \& Sutherland 1975; Cheng \& Ruderman 1980;
Ochelkov \& Usov 1984). Besides,
the radio emission of pulsars may be generated directly in the process
of the instability development (Asseo, Pellat \& Rosado 1980; Asseo,
Pellat \& Sol 1983).
Before to discuss the two-stream instability in the pulsar plasma
and the generation of coherent radio emission of pulsars,
we review both the physical processes that are responsible for
the filling of pulsar magnetospheres with plasma and the 
plasma parameters. 

\section{Physical processes and parameters of plasma in pulsar
magnetospheres}

A common point of all acceptable models of pulsars is that the 
spin-down power of strongly magnetized neutron stars is the energy 
source of non-thermal emission of pulsars. The rotational energy of 
the neutron star is transformed into the pulsar emission by
a long sequence of processes. 
%(e.g., Usov 1996).

\subsection{Physical processes in pulsar magnetospheres}

Strong electric fields are generated in the magnetospheres
of rotating magnetized neutron stars (for a review, see Michel 1991).
The component of the electric field ${\bf E}_\parallel =({\bf E\cdot B})
{\bf B}/|{\bf B}|^2$ along the magnetic field ${\bf B}$ is non-zero.
Primary particles are accelerated by this electric field
to ultrarelativistic energies and generate $\gamma$-rays. 
Some of these $\gamma$-rays are absorbed by creating
secondary $e^+e^-$ pairs. The created pairs
screen the electric field ${\bf E}_\parallel$ in the pulsar
magnetosphere everywhere except for compact regions.
The compact regions where ${\bf E}_\parallel$ is unscreened are
called gaps. These gaps are "engines" that are
responsible for the non-thermal radiation of pulsars. 
The gaps are located either near the magnetic poles of pulsars 
(polar gaps) or near their light cylinders (outer gaps).
Outer gaps may act as a generator of non-thermal radiation of
pulsars only if the period of the pulsar rotation is small enough, 
$P<P_{\rm cr}
\simeq {\rm a~few}\times 0.1$ s. For typical pulsars with 
$P>P_{\rm cr}$, the polar gap model has no an alternative.
In the polar gap model the sequence of processes that leads to 
the generation of non-thermal emission of pulsars is the 
following:

\begin{figure*}
\centerline{\psfig{file=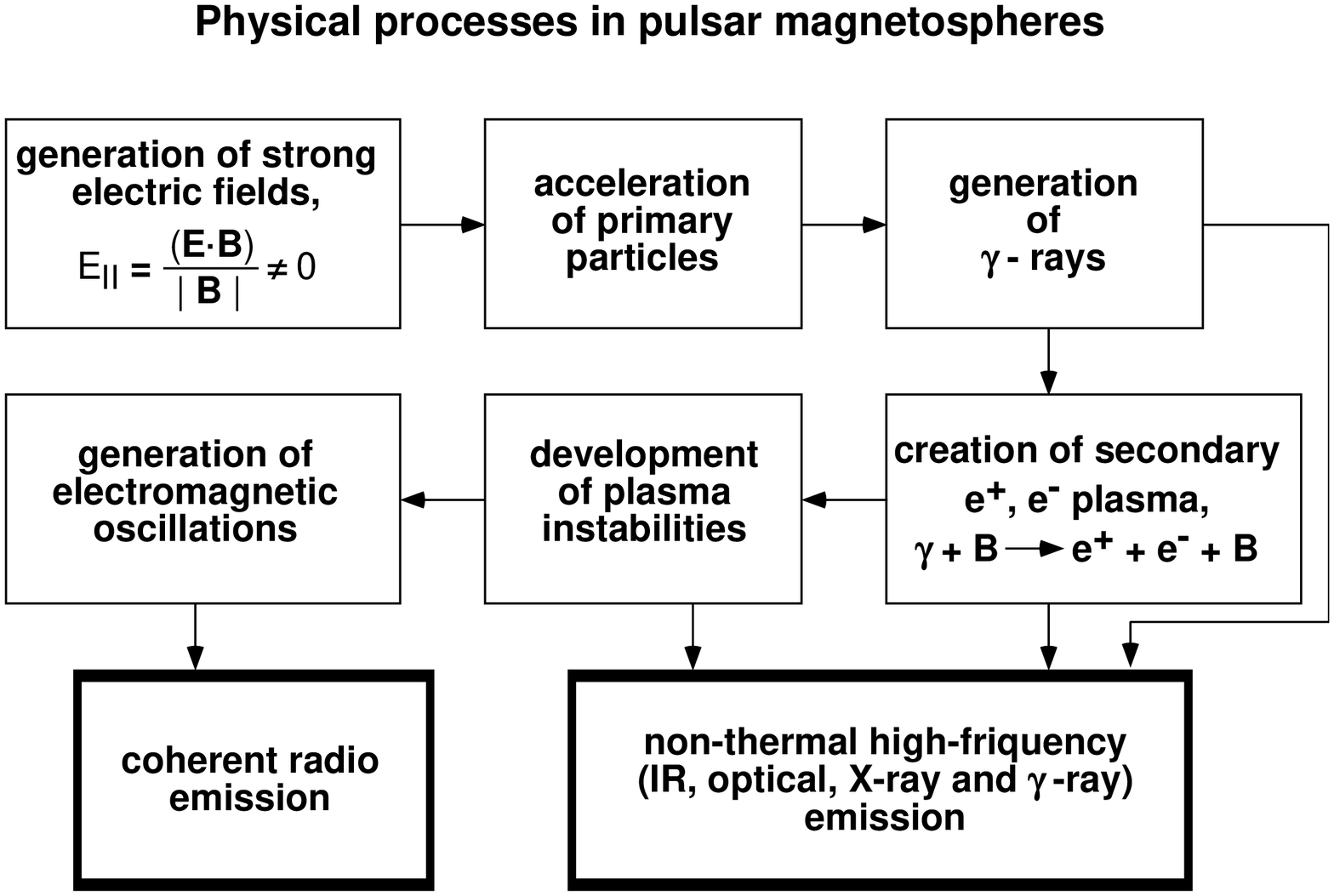,width=14cm,clip=} }
%\centerline{\psfig{file=figure1.eps,width=14cm,clip=} }
%\caption{Example of an included figure which covers two columns ( .ps file).
%\label{image}}
\end{figure*}

The density of primary particles accelerated in the polar gap 
is about the Goldreich-Julian density $n_{\rm GJ}=
|{\bf \Omega\cdot B}|/2\pi ce$, where ${\bf \Omega}$ is the angular velocity
of the pulsar rotation. The value of $n_{\rm GJ}$ is determined so
that the electric field $E_\parallel$ in the outflowing plasma is
screened completely if the charge density is equal to 
$en_{\rm GJ}$.
Therefore, the accelerating field $E_\parallel$ arises from deviations
from this density. Such a deviation may be because of (1) the inertia of
particles (Michel 1974), (2) the curvature of the magnetic field lines
(Arons 1981), the General Relativity effects (Muslimov \&
Tsygan 1992, Beskin 1992), and the binding of particles within the 
solid surface of a strongly magnetized neutron star (Ruderman \& 
Sutherland 1975; Cheng \& Ruderman 1980; Usov \& Melrose 1995, 1996).
When both the neutron star surface is cold enough and the surface
magnetic field is strong enough, there is no emission of particles
from the neutron star surface, and the binding of particles is 
responsible for the $E_\parallel$ field in the polar gaps
(e.g., Usov \& Melrose 1996).
In the case when particles flow freely from the stellar surface,
the field $E_\parallel$ is mainly due to the General
Relativity effects (i.g., Harding \& Muslimov 1998). 
The second case is applicable to typical pulsars
while the former case may be applied to pulsars with very strong
magnetic fields at their surface, $B_{_{\rm S}}\sim 10^{13}$ G
or more.
 
Ultrarelativistic primary particles in the process of their outflow
move practically along the magnetic field lines and
generate $\gamma$-rays via both curvature radiation and magnetic
Compton scattering. These $\gamma$-rays are absorbed in strong
magnetic fields of pulsar magnetospheres and create $e^+e^-$
pairs. Created pairs very quickly lose the
momentum component transverse to the magnetic field due to
synchrotron losses, and their distribution becomes one-dimensional,
i.e., ${\bf v}\parallel {\bf B}$. Hence, to develop the theory of
coherent radio emission of pulsars it is necessary to investigate both
the parameters of ultrarelatistic one-dimensional plasma outflowing
from the polar cap regions and instabilities of this plasma.

\subsection{Parameters of pulsar plasma} 

Electrons and positrons that
flow away from the neutron star vicinity with relativistic speeds may be
divided into two components: a secondary $e^+e^-$ plasma and an 
extremely high-energy beam of primary particles (see Fig.~1). 
Particles of the beam may 
be either electrons or positrons, depending on the direction of the 
electric field ${\bf E}_\parallel$ in the polar gap.
For the $e^+e^-$ plasma (denoted by $p$) and the ultrarelativistic beam
(denoted by $b$) the typical values of the Lorentz factor $\Gamma$
and the density $n$ are, respectively, (Ruderman \& Sutherland 1975;
Arons 1981a,b, 1984; Machabeli \& Usov 1989; Michel 1991)

\begin{equation}
\Gamma_p\simeq (2-3)R_c/R\simeq {\rm a~few} \times (1-10^2)\,,
\end{equation}

\begin{equation}
\Gamma_b\simeq e \Delta\varphi /m_ec^2\simeq {\rm a~few}\times 10^6\,,
\end{equation}

\begin{equation}
n_p\simeq (\Gamma_b/2\Gamma_p )n_b\,,
\end{equation}

\begin{equation}
n_b\simeq n_{_{\rm GJ}}=
\Omega B/(2\pi ce)\simeq 10^{11}(R/r)^3\,\,{\rm cm}^{-3}\,,
\end{equation}

\noindent
We have adopted here the following characteristic values for 
the pulsar parameters: $\Omega \simeq 10$ s$^{-1}$;
$B\simeq 10^{12}(R/r)^3$~G is the magnetic field in 
the magnetosphere; $R_c$ is the curvature radius of the magnetic field 
lines in the polar cap; 
$R\simeq 10^6$ cm is the neutron star radius; and
$\Delta\varphi$ is the potential drop across the polar gap.

\begin{figure*}
\centerline{\psfig{file=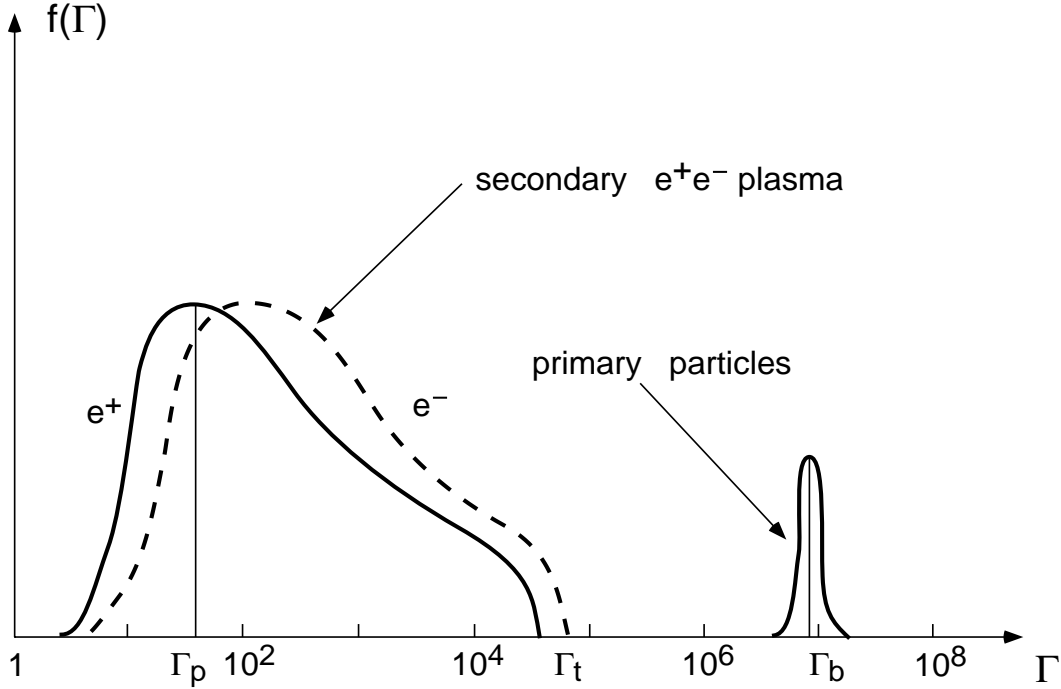,width=14cm,clip=} }
%\centerline{\psfig{file=figure1.eps,width=14cm,clip=} }
\caption{The energy spectrum of the basic components of the plasma
particles ejected from the polar gap region into the pulsar magnetosphere.
The spectra of secondary electrons (dashed line) and 
secondary positrons (solid line) are shifted relatively each other.
\label{image}}
\end{figure*}

\subsection{Nonstationarity of the plasma outflow}

It was suggested long time ago that the 
process of pair creation near the pulsar surface
is strongly nonstationary (Sturrock 1971; Alber et al. 1975;
Ruderman and Sutherland 1975). Such a nonstationarity was discussed
in detail in the outstanding paper by Ruderman and Sutherland (1975)
where the first self-consistent model of pulsar radiation
was presented. In the
Ruderman-Sutherland model, the creation of pairs occurs
in narrow spark discharges, separated from one another by a distance
of the order of the polar gap height $H$ (characteristic value of 
$H$ is $\sim 10^4$ cm). The secondary $e^+e^-$ plasma, produced by
absorption of $\gamma$-rays in a spark discharge, gathers into
an outflowing jet that has the discharge at its base.
Such a discharge is strongly nonstationary on a time scale of
$\sim 30(H/c)$ (Ruderman \& Sutherland 1975;
Cheng \& Ruderman 1980). The discharge nonstationarity results in
a strong modulation of the outflowing pair plasma along the jet. 
In fact, these jets consist of plasma clouds spaced by $L_1\simeq
30H\simeq 0.3R$.

However, it is not only the nonstationarity of discharges that can lead 
to inhomogeneity of the outflowing plasma along the fixed magnetic field
lines. Another reason  for such a inhomogeneity is the displacement 
of discharges caused by the curvature of the magnetic field lines
(Ruderman \& Sutherland 1975).
The displacement is directed toward the magnetic axis of the pulsar.
The speed with which a spark discharge moves across the polar cap is 
estimated (Ruderman \& Sutherland 1975; Cheng \& Ruderman 1980) as
$v_d\simeq cH/2R_c$. As already noted above, the average distance
between two discharges is $\sim H$. During the time $\tau_d\simeq
H/v_d\simeq 2R_c /c$, a discharge is displaced by the distance
of $\sim H$ and takes a position close to the initial position that 
of a neighboring discharge, which during the same time is also 
displaced by $\sim H$ toward the pulsar magnetic axis. As a result of
these displacements of discharges, the plasma outflowing along any fixed 
magnetic field lines becomes modulated with a characteristic length 
$L_2\simeq c\tau_d\simeq 2R_c$. Since, according to Ruderman and
Sutherland (1975) and Beskin (1982), in the polar gap the discharges 
are separated by intervals within which the process of pair creation
is suppressed, the plasma density in between the clouds falls down
almost to zero. Therefore, the outflowing plasma gathers into separate
clouds spaced by $L_2\simeq 2R_c$ along the direction of the plasma
outflow. In turn, these large clouds  
consist of small clouds with the lengths and the spaces
between them about $L_1\simeq 0.3R$. 
 
The idea that the process of pair creation near the polar caps
of pulsars is nonstationary has got recently the strong support 
(e.g., Muslimov \& Harding 1997; Harding \& Muslimov 1998). In these 
papers, the model of the polar gaps of typical pulsars which is 
rather close to reality is developed. In this model, it was shown
that both stable acceleration of primary particles and stable
creation of pairs are not possible. The characteristic length of 
the outflowing plasma modulation is expected more or less
comparable with $L_1$.

\section{Possible instabilities in pulsar plasma}

Development of instabilities in a magnetized plasma when
transverse electromagnetic waves ($t$-waves) are generated 
depends on the ratio $\omega_p/\omega_B$, where 

\begin{equation}
\omega_p=\sqrt{4\pi e^2n_p\over m_e}\simeq 5.64\times 10^4
n^{1/2}_p\,\,{\rm s}^{-1}\,,
\end{equation}

\begin{equation}
\omega_B={eB\over m_e c}\simeq 1.76\times 10^7B\,\,{\rm s}^{-1}\,,
\end{equation}

\noindent
are the Langmuir frequency of the plasma and the cyclotron frequency,
respectively. At the pulsar surface for the typical parameters of both 
the pulsar plasma and the strength of the magnetic field ($n_{p_0}\sim 
10^{14}~{\rm cm}^{-3}$ and $B_{_{{\rm S}_0}}\sim 10^{12}$~G), from equations
(5) and (6) we have $\omega_{p_0}/\omega_{_{{\rm B}_0}}\sim 10^{-7}$.

Since the difference between the phase velocity ($v_{\rm ph}$) of 
$t$-waves and the speed of light is extremely small at the pulsar 
surface (e.g., Lominadze, Machabeli \& Usov 1983),
 
\begin{equation}
c-v_{\rm ph}\sim c(\omega_{p_0}/\omega_{_{{\rm B}_0}})^2\Gamma_p\sim 
10^{-13}c
\end{equation}

\noindent 
in the frame of the pair plasma, particles of both the plasma and the 
beam cannot be in a resonance
with $t$-waves that propagate in the pulsar plasma, and resonance
instabilities (for example, the cyclotron instability) do not
develop in the neutron star vicinities.

Besides, the energy density of plasma ($\varepsilon_p\simeq m_ec^2
n_p\Gamma_p$) is very small in comparison with the energy density of
the magnetic field ($\varepsilon_{_{\rm B}}=B^2/8\pi$),

\begin{equation}
\varepsilon_p/\varepsilon_{_{\rm B}}\sim 2(\omega_{p_0}/
\omega_{_{{\rm B}_0}})^2\Gamma_p\sim 10^{-13}\,
\end{equation}

\noindent
Therefore, hydrodynamic instabilities are also suppressed near the pulsar 
surface by the strong magnetic field.

In the near zone of the pulsar magnetosphere ($r<c/\Omega$),
we have $n_p\propto B\propto r^{-3}$. 
The ratio $\omega_p/\omega_{_{\rm B}}$ increases with increase
of the distance from the pulsar, $\omega_p/\omega_{_{\rm B}}
\propto r^{3/2}$. Near the light cylinders of pulsars the ratio
$\omega_p/\omega_{_{\rm B}}$ is of the order of unity, and
the outflowing plasma may be unstable with respect to
excitation of $t$-waves (e.g., Machabeli \& Usov 1979, 1989;
Lominadze et al. 1983; Lyutikov, Machabeli \& Blandford 1999 and
references therein). At the distance from the pulsar in the 
range $R\leq r\ll c/\Omega$ only the two-stream instability
may be developed. 

Observations of pulsars suggest that the 
radio emission regions are far inside the light cylinder
(e.g., Gil \& Kijak 1993; Kramer et al. 1994; Kijak \& Gil 1997),

\begin{equation}
r_{\rm radio}\simeq 50R\simeq (0.02-0.04)\,c/\Omega\ll c/\Omega\,,
\end{equation}

\noindent
Therefore,
the development of the two-stream instability in the magnetospheres of 
pulsars is the most plausible reason of their extremely intense
radio emission.

\begin{figure}
\centerline{\psfig{file=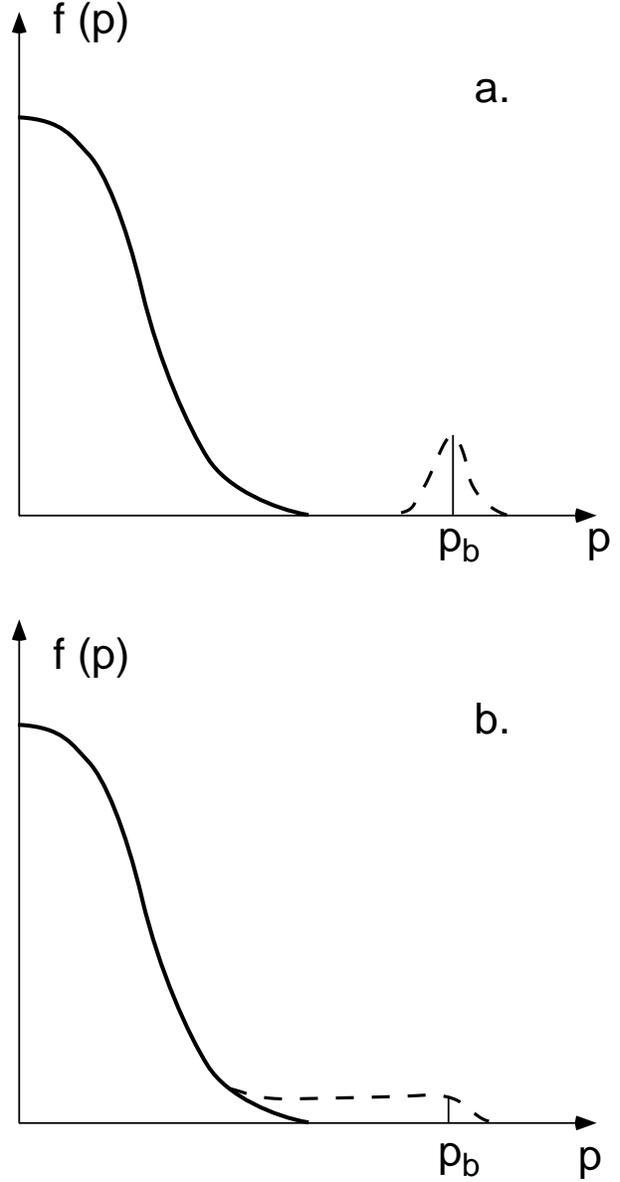,width=8cm,clip=} }
%\centerline{\psfig{file=figure1.eps,width=14cm,clip=} }
\caption{The momentum distribution function of particles
for a plasma-beam system: (a) prior and (b) after the 
development of the two-stream instability.
\label{image}}
\end{figure}

\section{Two-stream instability}

The two-stream instability may occur only if (1) there is
a "hump" on the momentum distribution function $f(p)$, 
i.e. a region having ${\partial f}/{\partial p} >0$, and
(2) in the frame where plasma is at rest the mean momentum of 
the "hump" particles ($p_{\rm b}$) is large enough in comparison 
with the mean momentum of the plasma particles (the Penrose
criteria). Particles, which are in the inside slope of the "hump"
where ${\partial f}/{\partial p} >0$, produce longitudinal ($l$)
waves with the frequency $\omega =({\bf k},{\bf v})$
(the Cherenkov resonance), where ${\bf k}$ is the wave vector 
(see Fig.~1). These particles thus lose their energy. As a result
the inside slope of the "hump" begins to move into the region
of even lower velocities, and the distribution function of the
"hump" particles becomes "plateau-shaped" (see, Fig.~1). The further
generation of $l$-waves is cut off. The plateau formation and the
cut-off in the plasma wave generation is called the quasi-linear
relaxation. The energy density of $l$-waves generated in the process
of quasi-linear relaxation is of the order of the energy density of 
the "hump" particles. 

A strong magnetic field has no an influence on the development of 
the two-stream instability in the one-dimensional plasma outflowing
from the polar cap regions of pulsars, and therefore, this instability 
may occur everywhere in the pulsar magnetosphere if the mentioned
above conditions for its development are realized.
Since there are several components of particles in the pulsar 
plasma that move with different speeds, the idea that 
the two-stream instability is responsible for the coherent radio 
emission of pulsars is very natural. 

The two-stream instability could be developed due to the following 
three reasons. The first one is the interaction between the
beam of primary particles and the secondary $e^+e^-$ plasma
(Ruderman \& Suthermand 1975). The second one is the interaction 
between electrons and positrons of the secondary plasma itself
(Cheng \& Ruderman 1977). At last, the two-stream
instability may be triggered in the inhomogeneous pulsar plasma
when the outflowing plasma clouds disperse and overlap each other
(Usov 1987; Ursov \& Usov 1988; Asseo \& Melikidze 1998).
 
\subsection{The interaction between the primary beam and the secondary
plasma}

The ultrarelativistic beam of primary particles has the distribution
function that obeys the conditions that are necessary for the two-stream
instability development (see Fig.~1). Therefore, the generation of radio 
emission of pulsars was initially associated with the ultrarelativistic
beam of primary particles (Ruderman \& Sutherland 1975). However,
further more precise estimates of the instability growth rate
have shown that it is impossible. Indeed, the characteristic time 
for development of the two-stream instability in the pulsar frame is 
(Benford \& Buschauer 1977; Egorenkov, Lominadze \& Mamradze 1983)

\begin{equation}
\tau_i\simeq (n_p/n_b)^{1/3}\Gamma_b\Gamma_p^{1/2}\omega_p^{-1}\,.
\end{equation}

From equations (1)-(4), (5) and (10), for typical pulsars the
value of $\tau_i$ is about $10^{-4}(r/R)^{3/2}$ s at the distance $r$
from the neutron star. For the instability development, it is necessary
that $\tau_i$ is, at least, a few times less than $\tau_0=r/c\simeq
3\times 10^{-5}(r/R)$ s, the characteristic
time of the plasma outflow from the distance $r$. However, the reverse 
relation between $\tau_i$ and $\tau_0$ holds. Hence, the two-stream
instability that, in principal, could be caused by the primary beam because 
its distribution function forms a "hump" with ${\partial f}/{\partial p} >0$,
in fact, has not enough time to be developed in the pulsar magnetospheres.
This is because both the beam density is too low and the Lorentz-factor of the 
beam is too high.

\begin{figure}
\centerline{\psfig{file=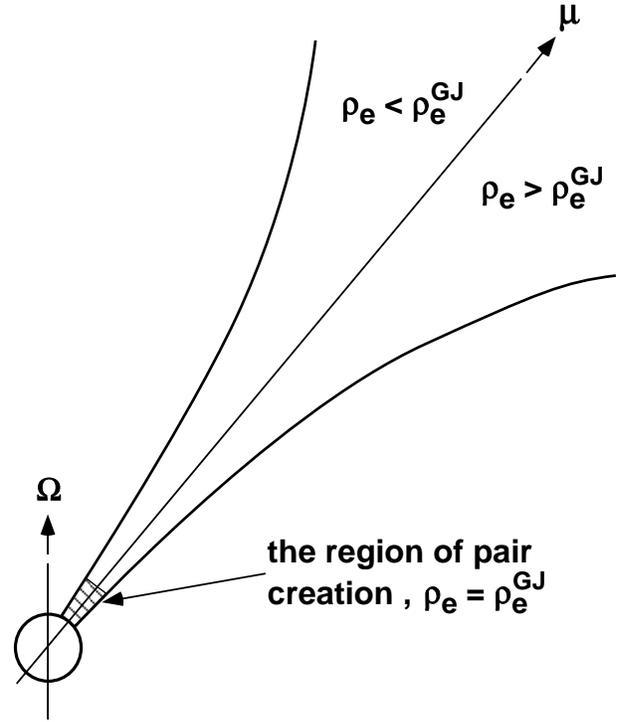,width=8cm,clip=} }
%\centerline{\psfig{file=figure1.eps,width=14cm,clip=} }
\caption{Charge flow separation of the pulsar plasma moving
outward along the curved magnetic field lines; ${\bf \mu}$
is the magnetic dipole moment.
\label{image}}
\end{figure}

\subsection{The interaction between the electrons and the positrons
of the secondary plasma}

The secondary $e^+e^-$ plasma produced by photons initially has identical 
distribution functions in its components (electrons and positrons).
In this case, the pair plasma is neutral, and the charge density 
$\rho_e$ is determined by the primary beam and equals to  
$\rho_e^{\rm GJ}=en_{\rm GJ}$.
In the process of the plasma outflow the charge density $\rho_e$ is
$\propto n\propto B$ while $\rho_e^{\rm GJ}$ is $\propto
B\cos\chi$, where $\chi$ is the angle between ${\bf \Omega}$
and ${\bf B}$. The magnetic field lines are curved, and therefore,
the angle $\chi$ varies with the distance from the pulsar.
This leads to a deviation of $\rho_e$ from
$\rho_e^{\rm GJ}$ and generation of the electric field along the 
magnetic field (see Fig.~3). This field accelerates one of the 
components of the secondary plasma and decelerates the other,
and the distribution functions of these components shift 
to each other.  
The difference between the mean velocities of the electrons and the
positrons is (Cheng \& Ruderman 1977)

\begin{equation}
|v_+-v_-|\simeq {n_b\over n_p}\left[{{\bf \Omega\cdot
B}\over ({\bf \Omega\cdot B})_0}-1\right]c\,,
\end{equation}

\noindent
where $({\bf \Omega\cdot B})_0$ is the ${\bf \Omega}$ and ${\bf B}$
product taken at the region of pair creation. 

\begin{figure*}
\centerline{\psfig{file=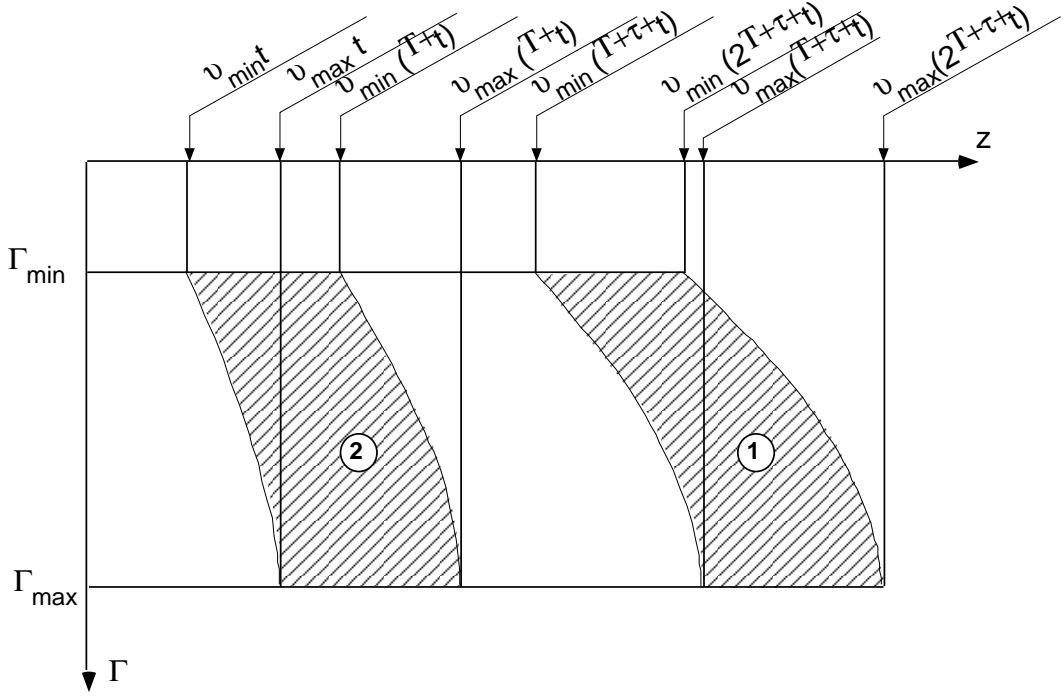,width=14cm,clip=} }
%\centerline{\psfig{file=figure1.eps,width=14cm,clip=} }
\caption{Particle distribution over the plane with coordinates
$z$ and $\Gamma$ at the instant $t$. The regions with particles are 
shaded. The first and the second clouds of plasma are numerated 
by 1 and 2.
\label{image}}
\end{figure*}

\noindent
Far the neutron star $(r\gg R)$, equations (3) and (11) yield

\begin{equation}
{|\Gamma_+-\Gamma_-|\over \Gamma_p}\simeq {r\Omega\over c}{\Gamma^3_p
\over \Gamma_b}
\end{equation}

\noindent
where 

\begin{equation}
|\Gamma_+-\Gamma_-|\simeq {|v_+- v_-|\over c}\Gamma_p^3\,.
\end{equation}

Using a relativistic generalization of the Penrose criteria 
(Buschauer \& Benford 1977; Krall \& Trivelpiece 1986) and
equation (12), the condition for development of the two-stream
instability may be written as 

\begin{equation}
\Gamma_p > 2\times 10^2(c/r\Omega)^{1/3}.
\end{equation}

\noindent 
From equation (14) we can see that the two-stream instability,
which develops because of the relative motion of the secondary plasma
components, may occur only rather near the light cylinder of pulsars.
Besides, the magnetic field near the pulsar surface must be close to 
the dipole field when the Lorentz-factor of the secondary
particles is equal to its maximum (see, equation (1)).

\subsection{The interaction between clouds of the secondary plasma}

Here the nonstationarity of plasma ejection from the polar gap regions
is modeled as follows. Plasma particles (electrons and positrons) are 
ejected along the $z$-axis from $z=0$ during the time $T$ with
Lorentz-factors ranging from $\Gamma_{\rm min}$ to $\Gamma_{\rm max}$,
so that $\Gamma_{\rm max}\gg\Gamma_{\rm min}\gg 1$. Then over the time 
$\tau$ no particles are ejected, after that everything is repeated.
In other words, we assume that the secondary $e^+e^-$ plasma
is ejected in the form of separate isolated clouds with the length
$cT$ along the $z$-axis, and the space between these clouds is 
$c\tau$.

At $z=0$, the distribution of ejected particles over Lorentz-factors
is described by the function $f_o(\Gamma )$ determined from the condition 
that the number of particles $dn_p$ per volume $dV$ with energies
from $m_ec^2\Gamma$ to $m_ec^2(\Gamma +d\Gamma )$ is equal to

\begin{equation}
dn_p=n_pdVf_o(\Gamma )d\Gamma\,.
\end{equation}

\noindent
The distribution function thus determined is normalized to unity,
i.e.,

\begin{equation}
\int_{\Gamma_{\rm min}}^{\Gamma_{\rm max}}f_o(\Gamma )d\Gamma =1\,.
\end{equation}

\noindent
The pulsar plasma can be regarded as collisionless. Therefore,
the spatial and energy distribution of outflowing particles at $z>0$ 
may be found from simple kinematic relations.

To study the dispersion of plasma clouds and their mutual overlapping
it is sufficient to consider the outflow of two successively
ejected clouds. It is expedient to draw the distribution of particles
in the $z$- and $\Gamma$-coordinates (see Fig.~4). In our case the
initial time ($t=0$) is the moment when the ejection of particles of
the second cloud is just finished. 

At the time

\begin{equation}
t_o={{v_{\rm min}(T+\tau )-v_{\rm max}T}\over {v_{\rm max} -
v_{\rm min}}}\simeq 2\Gamma^2_{\rm min}\tau
\end{equation}

\noindent
determined from the equation

\begin{equation}
v_{\rm min}(T+\tau +t_o)=v_{\rm max}(T+t_o)\,,
\end{equation}

\noindent
the fast particles of the second cloud catch up with the slow particles 
of the preceding cloud, and mutual overlapping of the clouds begins,
where

\begin{equation}
v_{\rm min}=[1-(1/\Gamma_{\rm min})^2]^{1/2}c\,,
\end{equation}

\begin{equation}
v_{\rm max}=[1-(1/\Gamma_{\rm max})^2]^{1/2}c\,.
\end{equation}

\begin{figure}
\centerline{\psfig{file=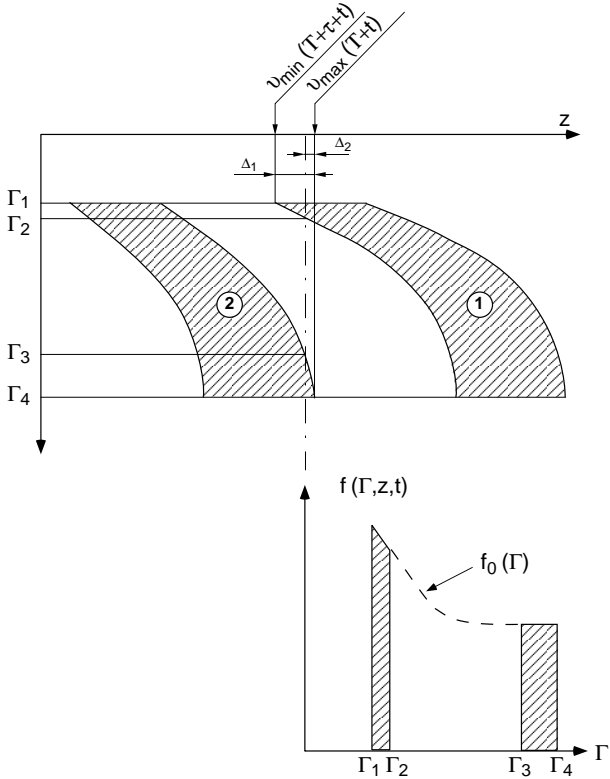,width=8cm,clip=} }
%\centerline{\psfig{file=figure1.eps,width=14cm,clip=} }
\caption{Overlapping of plasma clouds at $t>t_0$. 
The regions with particles are shaded. In the overlapping 
region, $z$ meets the condition $v_{\rm min}(T+\tau +t)<z<
v_{\rm max}(T+t)$. The distribution function of particles in 
the overlapping region at the distance $z$ from the pulsar surface 
is shown in the bottom. 
\label{image}}
\end{figure}

The width of the overlapping region at $t>t_o$ is

\begin{equation}
\Delta =v_{\rm max}(T+t)-v_{\rm min}(T+\tau +t)\,.
\end{equation}

The distribution function of particles in the overlapping region
is (see Fig.~5)

\begin{equation}
f(\Gamma ,z, t)=\cases{0\,,\,\,\,\,\,\,\,\,\,\,\,\,\,\,\,\,\Gamma
<\Gamma_1\,,\cr
f_o(\Gamma )\,,\,\,\,\,\,\Gamma_1<\Gamma<\Gamma_2\,,\cr
0\,,\,\,\,\,\,\,\,\,\,\,\,\,\,\,\,\,\Gamma_2<\Gamma_3\,,\cr
f_o(\Gamma )\,,\,\,\,\,\,\Gamma_3<\Gamma<\Gamma_4\,,\cr
0\,,\,\,\,\,\,\,\,\,\,\,\,\,\,\,\,\,\Gamma_4<\Gamma\,,\cr}
\end{equation}

\noindent
where

\begin{equation}
\Gamma_i=[1-(v_i/c)^2]^{-1/2},\,\,\,\,\,\,i=1,2,3,4,
\end{equation}

\begin{equation}
v_1=v_{\rm min},\,\,\,\,\,\,v_2={z\over T+\tau +t}\,,
\end{equation}

\begin{equation}
v_3={z\over T+t},\,\,\,\,\,\,v_4=v_{\rm max}.
\end{equation}

In the overlapping region, there are, in fact,
two streams of slow and fast particles (see Fig.~5).
This may lead to development of the two-stream instability.
A simple approximation where plasma is one-dimensional and
homogeneous may be employed for consideration of this instability
in the overlapping regions of plasma clouds in the magnetospheres 
of pulsars (Usov \& Ursov 1988). In this approximation, the 
dispersion equation for $l$-waves with ${\bf k}\parallel {\bf v}
\parallel {\rm B}$ is

\begin{equation}
1-{4\pi e^2n_p\over m_e}\int{f(\Gamma ,z,t )d\Gamma\over \Gamma^3
(\omega -kv)^2}=0
\end{equation}

At the initial stage of clouds overlapping, when the width of the 
overlapping region is much smaller than the width of the clouds,
along the $z$-axis, $\Delta\ll cT$, the function $f(\Gamma ,z,t)$
describes two groups of practically monoenergetic particles at 
velocities $v_2$ and $v_4$. Taking this into account, 
equation (26) may be written as

\begin{equation}
1-\omega_p^2\left[{\alpha\over (\omega-kv_2)^2}+
{\beta\over (\omega -kv_4)^2}\right]=0
\end{equation}

\noindent
where

\begin{equation}
\alpha=\int_{\Gamma_1}^{\Gamma_2}{f_o(\Gamma )\over \Gamma^3}d\Gamma\,,
\,\,\,\,\,\,\,
\beta =\int_{\Gamma_3}^{\Gamma_4}{f_o(\Gamma )\over \Gamma^3}d\Gamma\,.
\end{equation}

From equation (27), the maximum growth rate of $l$-waves is
(Ursov \& Usov 1988)

\begin{equation}
{\rm Im}\,\omega_{\rm max}={\sqrt{3}\over 2}\alpha^{1/2}
\left({\beta\over 2\alpha}\right)^{1/3}\omega_p\,,
\end{equation}

\noindent
and occurs for the wave vector

\begin{equation}
k_{\rm res}={\alpha^{1/2}\omega_p\over v_4-v_2}
\end{equation}

\noindent
that corresponds to a resonance between plasma oscillations of 
the slow ($\Gamma_1<\Gamma<\Gamma_2$) flow and high-energy particles
of the fast ($\Gamma_3<\Gamma<\Gamma_4$) flow.

Since for the slow and fast flows the $\Gamma$ ranges of plasma particles  
are narrow for slightly overlapping clouds ($\Delta\ll c\tau$), the
function $f_0(\Gamma )$ in equations (28) can be factored out the
integral sigh and then to integrate:

\begin{equation}
\alpha\simeq {f_o(\Gamma_1 )\over 2}
{\Gamma_2^2-\Gamma_1^2\over \Gamma_1^2\Gamma_2^2}=
{v_2-v_1\over c}f_o(\Gamma_1 )\simeq {\Delta_1\over ct_o}
f_o(\Gamma_{\rm min})
\end{equation}

\begin{equation}
\beta={f_0(\Gamma_4)\over 2}{\Gamma_4^2-\Gamma_3^2\over 
\Gamma_3^2\Gamma_4^2}=
{v_2-v_3\over c}f_o(\Gamma_4)\simeq {\Delta_2\over ct_o}
f_0(\Gamma_{\rm max}),
\end{equation}

\noindent
where (see Fig.~5)

\begin{equation}
\Delta_1=z-v_{\rm min}(T+\tau +t)\,,\,\,\,\,\,
\Delta_2=v_{\rm max}(T+t)-z
\end{equation}

\noindent
in the overlapping region, 

\begin{equation}
v_{\rm min}(T+\tau +t)<z<
v_{\rm max}(T+t)\,.
\end{equation}

Substituting equations (31) and (32) into equation (29), we
obtain

$$
{\rm Im}\,\omega_{\rm max}={\sqrt{3}\over 2^{4/3}}\omega_p
[f_o(\Gamma_{\rm min})]^{1/6}[f_o(\Gamma_{\rm max})]^{1/3}
$$

\begin{equation}
\,\,\times\left({\Delta -\Delta_2\over ct_o}\right)^{1/6}
\left({\Delta_2\over ct_o}\right)^{1/3}.
\end{equation}

The growth rate (35) depends on the distance $\Delta_2$ to the
leading front of the cloud overlapping and reaches the maximum

$$
({\rm Im}\,\omega_{\rm max})_{\rm max}={\omega_p\over 2}
[f_o(\Gamma_{\rm min})]^{1/6}[f_o(\Gamma_{\rm max})]^{1/3}
$$

\begin{equation}
\,\,\times ({\Delta/ ct_o})^{1/2}
\end{equation}

\noindent
at $\Delta_2=(2/3)\Delta$.

The distribution function of plasma particles in the
magnetospheres of pulsars may be roughly approximated
by the following law (Arons 1981)

\begin{equation}
f_o(\Gamma )=\cases{A\,\Gamma^{-3/2}\,\,\,\,\,\,\,\Gamma_{\rm min}
<\Gamma<\Gamma_*\,,\cr
A\,\Gamma^{-3/2}_*\,\,\,\,\,\,\,\Gamma_*<\Gamma<\Gamma_{\rm max}\,,\cr}
\end{equation}

\noindent
where $A$ is the constant determined from the normalization
condition (16) and written as

\begin{equation}
A={\Gamma_{\rm min}^{1/2}\over 2}\left[1+{1\over 2}\left(
{\Gamma_{\rm min}\over \Gamma_*}\right)^{1/2}
\left({\Gamma_{\rm max}\over \Gamma_*}-3
\right)\right]^{-1}.
\end{equation}

Substituting equation (37) into equation (36) and taking into 
account that $ct_o\simeq 2L\Gamma^2_{\rm min}$, we have the 
characteristic time of the instability development

\begin{equation}
\tau_i=({\rm Im}\,\omega_{\rm max})_{\rm max}^{-1}=
{2^{3/2}\Gamma^{5/4}_{\rm min}\Gamma_*^{1/2}\over \omega_p
A^{1/2}}\left({L\over \Delta}\right)^{1/2},
\end{equation} 

\noindent
where $L=c\tau$ is the space between the plasma clouds.

An instability manages to develop if the lifetime of the unstable
system exceeds the characteristic time of the instability development 
by more than an order. In our case, the unstable system is the
overlapping region, and its lifetime is $\Delta/c$. With a similar
criterion, $10\tau_i=\Delta_i/c$, we get the value of $\Delta$ at
which the two-stream instability is developing to be given by

\begin{equation}
\Delta_i\simeq (30c\omega_p^{-1}A^{-1/2}\Gamma_{\rm min}^{5/4}
\Gamma_*^{1/2}L^{1/2})^{2/3}.
\end{equation}

We now estimate the quantities that characterize the development of
the two-stream instability in the magnetospheres of pulsars.
The following parameters of the secondary $e^+e^-$ plasma may be
assumed: $\Gamma_{\rm min}\simeq 5$, $\Gamma_*\simeq 10^2$,
$\Gamma_{\rm max}\simeq 10^3$ (Arons 1981) and $L=c\tau\simeq
cT\simeq R\simeq
10^6$ cm (Usov 1987). From equations (3)-(5), (38) and (40),
we have $\Delta_i\simeq 0.8\times 10^5$ cm. Hence, the development of 
the two-stream instability already occurs when the cloud overlapping
is very small ($\Delta_i\ll L$). Further overlapping of the plasma clouds
($\Delta >\Delta_i$) could be considered only beyond the scheme of our 
linear approximation.

The distance from the neutron star to the region where the two-stream
instability occurs is

\begin{equation}
r_i\simeq 2c\tau\Gamma^2_{\rm min}=2L\Gamma^2_{\rm min}
\end{equation}

\noindent
or numerically $r_i\simeq 5\times 10^7$ cm for $L\times 10^6$ cm and
$\Gamma_{\rm min}\simeq 5$. The value of $r_i$ is consistent with
the distance from the neutron stars
to the radio emission regions (see equation (9)).

The existence of several scales of the outflowing plasma modulation 
(e.g., $L_1\simeq 0.3R$ and $L_2\simeq 2R_c$ in the 
Ruderman-Sutherland model) may lead to the situation that 
in the magnetospheres of pulsars there are several regions where
the two-stream instability develops and the generation of coherent 
radio emission occurs. From the observational data on radio emission of
pulsars it follows that such a situation is possible for
some pulsars (e.g., Gil 1985).

\section{Discussion}

The mechanism of coherent radio emission of pulsars is closely connected 
with the location of the radio emitting regions in the pulsar magnetospheres.  
If it is confirmed that the distances from the neutron stars to 
the radio emitting regions are $\ll c/\Omega$ (Gil \& Kijak 1993;
Rankin 1993; Kramer et al. 1994; Kijak \& Gil 1997), i.e.,
these regions are far inside the light cylinder, then most probably,
the two-stream instability developed in the outflowing strongly 
nonhomogeneous plasma (Usov 1987; Ursov \& Usov 1988; Asseo \& Melikidze 
1998) is responsible for the generation of the pulsar radio emission.
It is worth noting that if the process of pair creation near the pulsar
surface in strongly nonstationary indeed, the development of this instability 
in the magnetospheres of pulsars is almost inevitable.

Longitudinal (nonescaping) Langmuir waves are generated in the process of
development of the two-stream instability. Then, these waves may be
converted into $t$-waves, that can escape from the pulsar magnetosphere,
by means of the following nonlinear effects:

\begin{figure*}
\centerline{\psfig{file=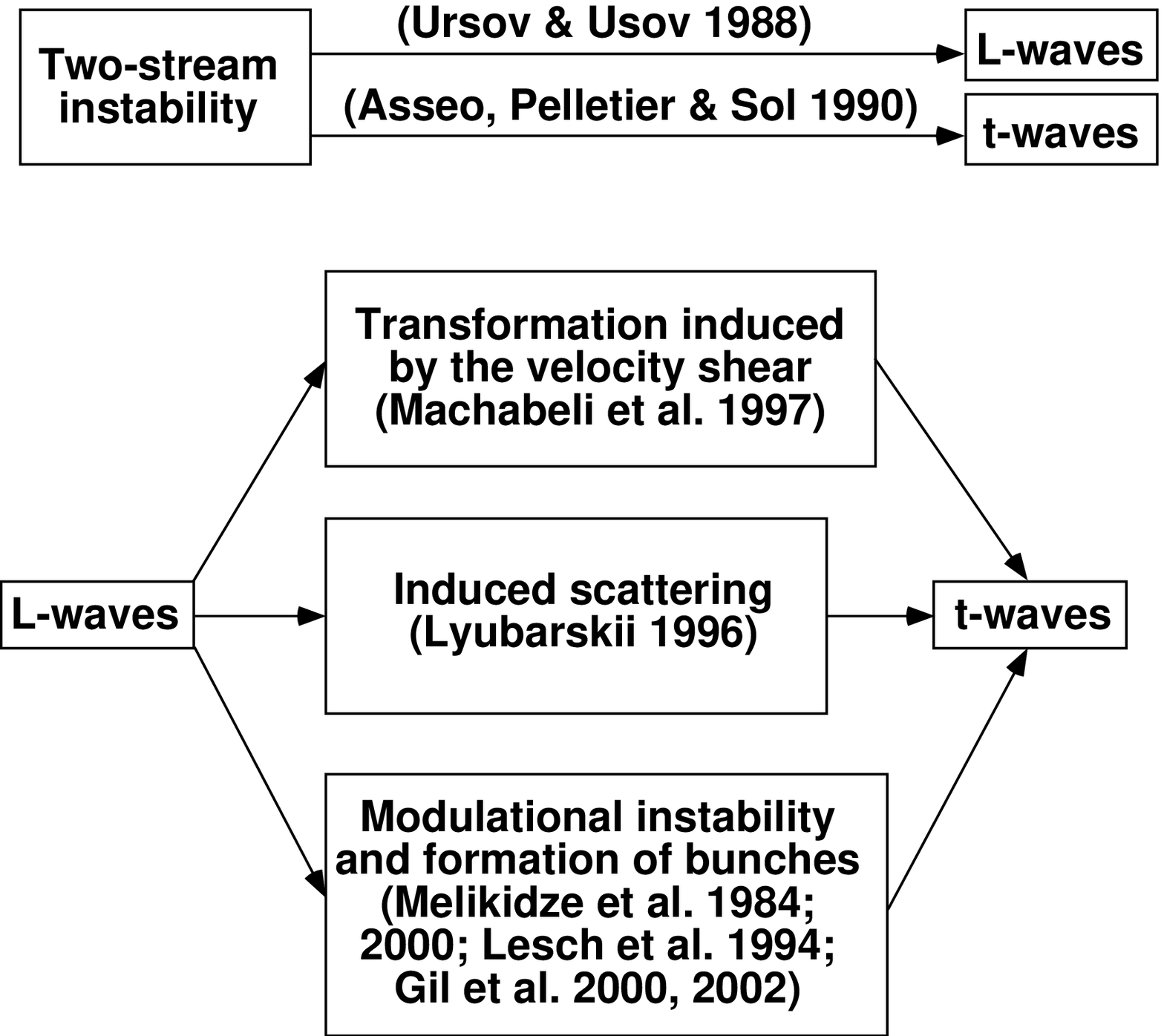,width=14cm,clip=} }
%\centerline{\psfig{file=figure1.eps,width=14cm,clip=} }
%\caption{
%\label{image}}
\end{figure*}

\vskip 0.4cm

\begin{acknowledgements}
I gratefully acknowledge the support by the Heraeus foundation.
This work was supported by MINERVA Foundation, Munich, Germany.
\end{acknowledgements}
   
% Example list of References


\begin{thebibliography}{} 
\bibitem{AKE75} Alber Ya.I., Krotova Z.N., Eidman V. Ya., 1975,
Astrophizika 11, 283
\bibitem{A81a} Arons J., 1981a, in Proc. Internat. Summer School and 
Workshop in Plasma Physics, ed. T.D. Guyenne (ESA SP-161), 273
\bibitem{A81b} Arons J., 1981b, in Proc. IAU Symp. 95, Pulsars,
eds. W. Sieber and R. Wielebinski, Reidel, Dordrecht, 69
\bibitem{A81c} Arons J., 1981c, ApJ 248, 1099
\bibitem{A84} Arons J., 1984, Adv. Space Res. 3, 287
\bibitem{A93} Asseo E., 1993, MNRAS 264, 940
\bibitem{A96} Asseo E., 1996, in Proc. Pulsars: Problems and
Progress, ASP Conf. Ser. 105, eds. S. Johnston, M.A. Walker and
M. Bailes, 147
\bibitem{AM98} Asseo E., Melikidze G.I., 1998, MNRAS 301, 59

\bibitem{APS90}Asseo E., Pelletier, G., Sol, H., 1990, MNRAS
247, 529 

\bibitem{APR80} Asseo E., Pellat R., Rosado M., 1980, ApJ 239, 661

\bibitem{APS83} Asseo E., Pellat R., Sol H., 1983, ApJ 266, 201


\bibitem{BB77a} Benford G., Buschauer R., 1977, MNRAS 179, 189

\bibitem{B82} Beskin V.S., 1982, Soviet Astron. 26, 443 

\bibitem{B92} Beskin V.S., 1992, Soviet Astron. Lett. 16, 286

\bibitem{BB77b} Buschauer R., Benford R., 1977, MNRAS 179, 99


\bibitem{CR77} Cheng A.F., Ruderman M.A. 1977, ApJ 212, 800

\bibitem{CR80} Cheng K.S., Ruderman M.A. 1980, ApJ 235, 576

\bibitem{ELM83} Egorenkov V.D., Lominadze J.G., Mamradze P.G.,
1983, Astrophizika 19, 753

\bibitem{G85} Gill J.A., 1985, ApJ 299, 154

\bibitem{GK93} Gil J.A., Kijak J., 1993, A\&A 273, 563

\bibitem{G02} Gil J.A. et al., 2002, these proceedings

\bibitem{HM98} Harding A.K., Muslimov A.G., 1998, ApJ 508, 328

\bibitem{KG97} Kijak J., Gil J.A., 1997, MNRAS 288, 631

\bibitem{KT86} Krall N.A., Trivelpiece A.W., 1986, Principles of 
Plasma Physics, San Francisco Press

\bibitem{K94} Kramer M., Wielebinski R., Jessner A., Gil J.A.,
Seiradakis J.H., 1994, A\&AS 107, 515

\bibitem{LGS94} Lesch H., Gil J.A., Shukla P.K., 1994, Space Sci. Rev.
68, 349

\bibitem{LMU83} Lominadze J.G., Machabeli G.Z., Usov V.V., 1983,
A\&SS 90, 19

\bibitem{L96} Lyubarsky Yu.E., 1996, A\&A, 308, 809

\bibitem{LMB99} Lyutikov M., Machabeli G.Z., Blandford R.D., MNRAS
305, 338

\bibitem{MMR97} Machabeli G.Z., Mahajan S.M., Rogova A.D., 1997, ApJ
478, L129

\bibitem{MU79} Machabeli G.Z., Usov V.V., 1979, Soviet Astron. Lett. 5, 238

\bibitem{MU89} Machabeli G.Z., Usov V.V., 1989, Soviet Astron. Lett. 15, 393 

\bibitem{MGP00} Melikidze G.I., Gil J.A., Pataraya A.D., 2000
ApJ 544, 1081

\bibitem{MP84} Melikidze G.I., Pataraya A.D., 1984, Astrophizika 20, 157


\bibitem{M81} Melrose D.B., 1981, in Proc. IAU Symp. 95, Pulsars,
eds. W. Sieber and R. Wielebinski, Reidel, Dordrecht, 133
\bibitem{M93} Melrose D.B., 1993, in Pulsars as Physics Laboratories,
eds. R.D. Blandford, A. Hewish and L. Mestel, Oxford University 
Press, 105
\bibitem{M95} Melrose D.B., 1995, J. Astrophys. Astr. 16, 137

\bibitem{M74} Michel F.C., 1974, ApJ 192, 713

\bibitem{M91} Michel F.C., 1991, Theory of Neutron Star 
Magnetospheres, UCP

\bibitem{MH97} Muslimov A.G., Harding A.K., 1997, ApJ 485, 735

\bibitem{MT92} Muslimov A.G., Tsygan A.I., 1992, MNRAS 255, 61


\bibitem{OU84} Ochelkov Yu.P., Usov V.V., 1984, Nature 309, 332

\bibitem{R93} Renkin J.M., 1993, ApJ 405, 285

\bibitem{RS75} Ruderman M.A., Sutherland P.G., 1975, ApJ 196, 51

\bibitem{S71} Sturrock P.A., 1971, ApJ 164, 529


\bibitem{UU88} Ursov V.N., Usov V.V., 1988, Ap\&SS 140, 325
\bibitem{U81} Usov V.V., 1981, Adv. Space Rec. 1, 125
\bibitem{U87} Usov V.V., 1987, ApJ 320, 333

\bibitem{U00} Usov V.V., 1996, in Proc. Pulsars: Problems and
Progress, ASP Conf. Ser. 105, eds. S. Johnston, M.A. Walker and
M. Bailes, 323

\bibitem{UM95} Usov V.V., Melrose D.B., 1995, Australian J. Phys. 48, 571

\bibitem{UM96} Usov V.V., Melrose D.B., 1996, ApJ 464, 306

 


\end{thebibliography}
\end{document}